\documentclass[
    ,final            % use final for the camera ready runs
  ]
  {aipproc}

\layoutstyle{8x11single}

\def\CHP {CH$^+$}
\def\THCHP {$^{13}$CH$^+$}
\def\OHP {OH$^+$}
\def\SHP {SH$^+$}
\def\NHP {NH$^+$}
\def\WAT {H$_2$O}
\def\WATP {H$_2$O$^+$}
\def\HTOP {H$_3$O$^+$}

\def\HTHCOP {H$^{13}$CO$^+$}
\def\CCCHH {C$_{3}$H$_2$}

\def\kms    {\ifmmode{{\rm km~s}^{-1}}\else{km~s$^{-1}$}\fi}
\def\mum     {\ifmmode{\mu{\rm m}}\else{$\mu{\rm m}$}\fi}
\def\cmsq  {$\hbox{{\rm cm}}^{-2}$}    %cm-2

\begin{document}

\title{APEX telescope observations of new molecular ions}

\classification{<Replace this text with PACS numbers; choose from this list:
                \texttt{http://www.aip..org/pacs/index.html}>}
%\keywords      {<Enter Keywords here>}

\author{F.~Wyrowski}{
  address={Max-Planck-Institut f\"ur Radioastronomie, Auf dem H\"ugel 69, 
           53121 Bonn, Germany }
}

\author{K.~M.~Menten}{
%  address={<common address for author2 and author3>}
}

\author{R.~G\"usten}{
%  address={<common address for author2 and author3>}
%  ,altaddress={<author1 address>} % additional visiting address
}
\author{A.~Belloche}{
%  address={<common address for author2 and author3>}
%  ,altaddress={<author1 address>} % additional visiting address
}

\begin{abstract}
  Hydrides are key ingredients of interstellar chemistry since they
  are the initial products of chemical networks that lead to the
  formation of more complex molecules. The fundamental rotational
  transitions of light hydrides fall into the submillimeter wavelength
  range. Using the APEX telescope, we observed the long sought
  hydrides \SHP\ and \OHP\ in absorption against the strong continuum
  source Sagittarius B2(M). Both, absorption from Galactic center gas
  as well as from diffuse clouds in intervening spiral arms
  over a large velocity range is observed. The detected absorption of a
  continuous velocity range on the line of sight shows these hydrides
  to be an abundant component of diffuse clouds. In addition, we used
  the strongest submillimeter dust continuum sources in the inner
  Galaxy to serve as background candles for a systematic census of
  these hydrides in diffuse clouds and massive star forming regions of
  our Galaxy and initial results of this survey are presented.
\end{abstract}

\maketitle

%%%%%%%%%%%%%%%%%%%%%%%%%%%%%%%%%%%%%%%%%%%%
%% MAINMATTER
%%%%%%%%%%%%%%%%%%%%%%%%%%%%%%%%%%%%%%%%%%%%

\section{Interstellar hydrides}

Hydrides were the first molecules detected in interstellar space.
More then 70 years ago, optical absorption lines from CH and \CHP\
were detected in stellar spectra by \citet{Dunham1937} due to
intervening interstellar diffuse clouds on the lines-of-sight. Diffuse
clouds are relatively tenuous, exposed to the UV radiation from stars.
A comprehensive review of diffuse clouds and the molecular detections
in them is given in \citet{SnowMcCall2006}.  Hydrides are key
ingredients of interstellar chemistry since they are the initial
products of chemical networks that lead to the formation of more
complex molecules.  The fundamental rotational transitions of light
hydrides fall into the submillimeter wavelength range, among them also
several transitions of the water molecule. As an abundant component of
the earth's atmosphere, water is responsible for blocking large
portion of the submillimeter wavelengths range for ground-based
observations. Only from exceptional sites, such as the the dry high
altitude Chajnantor plateau in the Atacama desert of northern Chile,
reasonable transparency in several atmospheric windows is reached over
substantial time periods that is exploited by the APEX, ASTE and
NANTEN2 single dish telescopes and soon the Atacama Large Millimeter
Array interferometer (ALMA).  For a review of hydride discoveries using
the Caltech Submillimeter Observatory (CSO) on Mauna Kea see
\citet{Lis2009}. Here we summarize recent new observations of
interstellar hydrides using the Atacama Pathfinder EXperiment, APEX,
see \citet{Gusten_etal2006}, that complement the flood of new hydride
results obtained with the Herschel Space Observatory.

\section{New molecular ions towards Sgr B2 (M)}

Sagittarius B2 is the most massive star forming region in our
Galaxy. It harbors the most luminous hot molecular cores, Sgr B2(M)
and (N), and is situated close to the Galactic Center at a distance of
about 8~kpc. The high gas column density, high excitation and rich
chemistry make this region the best hunting ground for new
interstellar molecules and many of the molecules were indeed firstly
discovered there. In addition, the line of sight towards the source
crosses several galactic spiral arms, which enables absorption studies
of intervening diffuse clouds. The strong continuum emission of Sgr B2
is due to luminous ultracompact HII regions at centimeter wavelengths
and thermal dust continuum emission from their massive envelopes in the
submillimeter/far-infrared.

\begin{figure}[th]
  \includegraphics[width=0.65\textwidth,angle=-90]{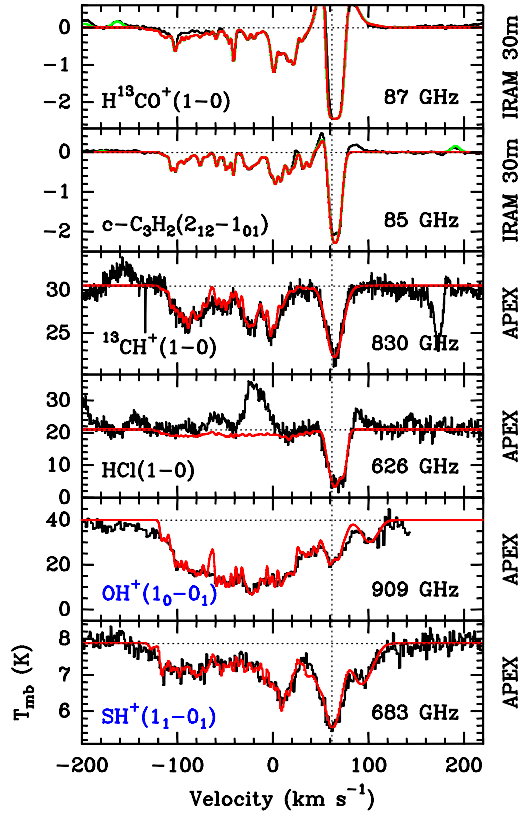}
  \caption{
New hydrides observations with APEX towards Sgr B2(M) (\citet{menten+2010}, 
\citet{Wyrowski2010a}; the upper two panels 
show IRAM 30m detections for comparison).
}
\end{figure}

The 12m APEX telescope, a modified copy of the US ALMA prototype antenna,
was used in conjunction with the powerful CHAMP+ heterodyne array
described in \citet{Kasemann2006} to observe toward Sgr B2(M) several
hydrides in the 450 and 350~\mum\ atmospheric windows, two of them for
the first time in interstellar space. Figure 1 shows the measured
spectra of \THCHP, HCl, \OHP\ and \SHP\ together with observations of
\HTHCOP\ and \CCCHH\ taken from a 3mm line survey carried out with the
IRAM 30m telescope for comparison (see \citet{belloche2008}).  Besides
strong absorption at the source velocity around 65~\kms, all molecules
except HCl show in addition absorption from a large range of
velocities from diffuse clouds on the line of sight. This demonstrates
that the newly detected molecular ions \SHP\ and \OHP\ are an abundant
component of diffuse clouds (\citet{menten+2010},
\citet{Wyrowski2010a}).

At the velocity of the Sgr B2(M) cloud, the \SHP\ column density is
$2\times 10^{14}$~\cmsq\ and in the diffuse line-of-sight clouds
column densities per unit velocity interval of 0.5 to $6 \times
10^{12}$~\cmsq/(\kms) are found.  \THCHP\ can be used together with
the Galactic gradient of the $^{12}$C/$^{13}$C ratio measured by
\citet{Milam05} to determine the column densities of \CHP. The
SH$^+$/$^{12}$CH$^+$ column density ratio varies significantly between
velocity components, from $0.07$ to 1.2.

%OH+
An \OHP\ column density local to Sgr B2(M) of $2.6 \times 10^{14}$~\cmsq\ is
found. On the intervening line of sight, the column density per unit
velocity interval is in the range of 1 to $40 \times
10^{12}$~\cmsq/(\kms). \OHP\ is found to be on average more abundant
than the other hydrides \SHP\ and \CHP. Abundance ratios of OH
and atomic oxygen to \OHP\ are found in the range of $10^{1-2}$ and
$10^{3-4}$, respectively.

% TODO:
% hcl

\section{Inner galaxy survey of \OHP}

\begin{figure}[th]
  \includegraphics[width=0.6\textwidth,angle=-90]{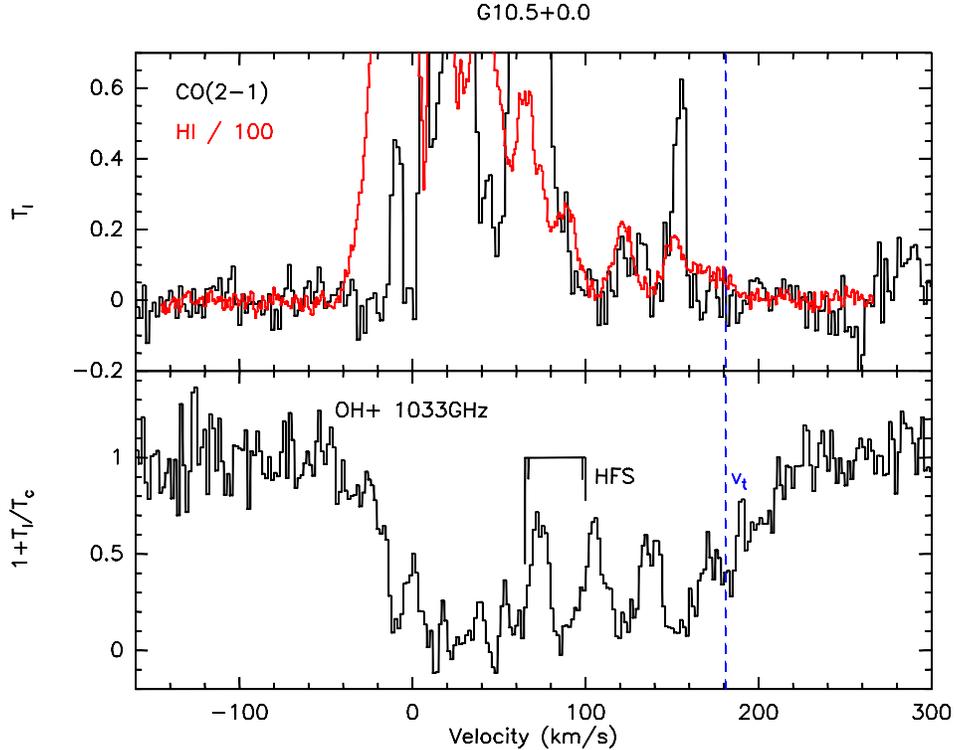}
  \caption{
APEX THz observations of OH$^+$ towards the massive star forming clump 
G10.5+0.0 (lower panel), compared with CO(2--1) (APEX) and HI 
(divided by 100, extracted from the Southern Galactic Plane Survey, \citet{McClure+2005}). The hyperfine structure pattern of OH$^+$ as well as the
the tangential velocity at the source galactic longitude are indicated.
}
\end{figure}

The rotational ground state transition of \OHP\ is split into three
lines at 909, 971 and 1033~GHz, each of them showing further hyperfine
splitting. Besides the 909~GHz lines, also the 1033~GHz line can be
observed from the ground from an excellent site such as the Chajnantor
plateau in the first atmospheric THz window. This was done for the
first time in June 2010 with the APEX telescope using the new MPIfR
THz receiver (Wyrowski et al., Leinz et al., both in prep.). Since the 1033~GHz
lines are stronger than the 909~GHz lines, they allow conducting a
survey of \OHP\ towards strong continuum sources in a reasonable
amount of time. Twenty-five of the strongest submillimeter continuum sources,
all luminous massive star forming regions, were observed in the inner
Galaxy during excellent atmospheric conditions with precipitable
water vapor of about 0.2--0.3~mm, resulting in atmospheric
transmission of up to 30\%. \OHP\ could be detected towards most of
the sources as a further confirmation for the widespread distribution
of \OHP\ in the interstellar medium.  Figure 2 shows as a preliminary
result the deep absorption found towards the massive star forming
clump G10.47+0.03, which is much broader than the maximal separation
of hyperfine components of about 35~\kms. Comparing the absorption
with emission of CO and HI towards this source, it is found that all of
the \OHP\ absorption components have HI counterparts, while in a few
\OHP\ velocity ranges, no CO was detected. This is especially evident
at the extreme velocities close to the tangential velocity on this
line of sight. Since the velocity of the star forming region itself is
at about 60~\kms, the detection of velocity components beyond the
source velocity shows that the region is situated at the far kinematic
distance at a distance of 10.6~kpc, which is consistent with the HI
absorption results by \citet{pandian+2008}. This result shows that
\OHP\ absorption studies might be employed to break the near/far
kinematic distance ambiguity for sources that show only strong
submillimeter continuum and no centimeter continuum needed for HI
absorption studies.

\section{Prospects for ALMA}

The ground-based hydride observations with the CSO (\citet{Lis2009})
and APEX pave the way for future studies with high angular resolution
and sensitivity using the ALMA interferometer, currently being
constructed on the Chajnantor plateau, with early science likely
starting already late in 2011. Table 1 summarizes important hydride lines
that will be observable with ALMA. For sources with considerable
redshift even more hydride transitions come into reach, e.g. ground
state lines from \CHP, \WAT, \HTOP, HF and \NHP, although the latter
has so far not been detected in interstellar space. The recent
Herschel extragalactic detections of \WATP\ in nearby starburst
galaxies (\citet{weiss}, \citet{vdwerf}) offer promising
prospects for high redshift detections. The usual excellent bandpass
stability of interferometers will facilitate those detections.
ALMA will be able to map the absorption in front of the spatially
extended submillimeter dust continuum. This will allow to study the
small scale structure of the hydrides in the diffuse clouds in
conjunction with high velocity resolution of the lines. ALMA will
also allow detailed mapping of hydrides in the outflows from young
stellar objects to shed light onto the detailed chemical processes
in the flows.

\begin{table}[ht]
\begin{tabular}{lclrl}
\hline
Molecule & Transition & Frequency & Band & Reference/Note \\
         &            &  (GHz)    &      &        \\
\hline
OH$^+$ & 1--0         & 909.1588  & 10 & \citet{Wyrowski2010a} \\
SH$^+$ & 1--0         & 683.4223  &  9 & \citet{menten+2010} \\
SH$^+$ & 1--0         & 345.9298  &  7 & close to CO \\
SH$^+$ & 2--1         & 893.1338  & 10 & \citet{menten+2010} \\
$^{13}$CH$^+$ & 1--0  & 830.2161  & 10 & \citet{menten+2010} \\
HCl    & 1--0         & 625.9188  & 9 & \citet{menten+2010} \\
H$_2$Cl$^+$ & 1--0    & 485.4208 & 8 & \citet{Lis2010b} \\  
%H$_2$Cl$^+$ & 2--1    & 781.6268 & & \citet{Lis2010b} \\  
H$_2$O$^+$ & 3/2--1/2 & 604.6786 & 9 & \citet{Schilke+2010} \\
H$_2$O$^+$ & 3/2--3/2 & 607.2273 & 9 & \citet{Schilke+2010} \\
H$_2$O$^+$ & 1/2--1/2 & 631.7241 & 9 & \citet{Schilke+2010} \\
H$_2$O$^+$ & 1/2--3/2 & 634.2729 & 9 & \citet{Schilke+2010} \\
NH         & 1--0     & 946.4758 & 10 & \citet{hilyblant2010} \\
NH$_2$ & 3/2--3/2         & 462.4335 & 8 & \citet{vanDishoeck1993} \\
\hline
\end{tabular}
\caption{Rotational lines of hydrides accessible with ALMA
         from the ground. In case of hyperfine structure, only
         the strongest component is given. References to
         astronomical observations of the lines are given.}
\label{tab:a}
\end{table}

%%%%%%%%%%%%%%%%%%%%%%%%%%%%%%%%%%%%%%%%%%%%%%%%
%% BACKMATTER
%%%%%%%%%%%%%%%%%%%%%%%%%%%%%%%%%%%%%%%%%%%%%%%%

%\begin{theacknowledgments}
%\end{theacknowledgments}



\bibliographystyle{aipproc}   % if natbib is available

\bibliography{wyrowski-smiles2010.bib}

\begin{thebibliography}{17}
\expandafter\ifx\csname natexlab\endcsname\relax\def\natexlab#1{#1}\fi
\providecommand{\enquote}[1]{``#1''}
\expandafter\ifx\csname url\endcsname\relax
  \def\url#1{\texttt{#1}}\fi
\expandafter\ifx\csname urlprefix\endcsname\relax\def\urlprefix{URL }\fi
\providecommand{\eprint}[2][]{\url{#2}}

\bibitem[{Dunham}(1937)]{Dunham1937}
T.~{Dunham}, \emph{\pasp} \textbf{49}, 26--28 (1937).

\bibitem[{Snow} and {McCall}(2006)]{SnowMcCall2006}
T.~P. {Snow}, and B.~J. {McCall}, \emph{\araa} \textbf{44}, 367--414 (2006).

\bibitem[{Lis} et~al.(2009)]{Lis2009}
D.~C. {Lis}, P.~F. {Goldsmith}, E.~A. {Bergin}, E.~{Falgarone}, M.~{Gerin}, and
  E.~{Roueff}, \enquote{{Hydrides in Space: Past, Present, and Future},} in
  \emph{Astronomical Society of the Pacific Conference Series}, edited by
  {D.~C.~Lis, J.~E.~Vaillancourt, P.~F.~Goldsmith, T.~A.~Bell, N.~Z.~Scoville,
  \& J.~Zmuidzinas}, 2009, vol. 417 of \emph{Astronomical Society of the
  Pacific Conference Series}, pp. 23--24.

\bibitem[{G{\"u}sten} et~al.(2006)]{Gusten_etal2006}
R.~{G{\"u}sten}, L.~{\AA}. {Nyman}, P.~{Schilke}, K.~{Menten}, C.~{Cesarsky},
  and R.~{Booth}, \emph{\aap} \textbf{454}, L13--L16 (2006).

\bibitem[{Menten} et~al.(2010)]{menten+2010}
K.~{Menten}, F.~{Wyrowski}, A.~{Belloche}, R.~{G\"usten}, H.~{M\"uller}, and
  L.~{Dedes}, \emph{\aap\ in press, preprint arXiv:1009.2825}  (2010).

\bibitem[{Wyrowski} et~al.(2010)]{Wyrowski2010a}
F.~{Wyrowski}, K.~M. {Menten}, R.~{G{\"u}sten}, and A.~{Belloche}, \emph{\aap}
  \textbf{518}, A26+ (2010).

\bibitem[{Kasemann} et~al.(2006)]{Kasemann2006}
C.~{Kasemann}, R.~{G{\"u}sten}, S.~{Heyminck}, B.~{Klein}, T.~{Klein}, S.~D.
  {Philipp}, A.~{Korn}, G.~{Schneider}, A.~{Henseler}, A.~{Baryshev}, and T.~M.
  {Klapwijk}, \enquote{{CHAMP+: a powerful array receiver for APEX},} in
  \emph{Society of Photo-Optical Instrumentation Engineers (SPIE) Conference
  Series}, 2006, vol. 6275 of \emph{Society of Photo-Optical Instrumentation
  Engineers (SPIE) Conference Series}.

\bibitem[{Belloche} et~al.(2008)]{belloche2008}
A.~{Belloche}, K.~M. {Menten}, C.~{Comito}, H.~S.~P. {M{\"u}ller},
  P.~{Schilke}, J.~{Ott}, S.~{Thorwirth}, and C.~{Hieret}, \emph{\aap}
  \textbf{482}, 179--196 (2008), \eprint{0801.3219}.

\bibitem[{Milam} et~al.(2005)]{Milam05}
S.~N. {Milam}, C.~{Savage}, M.~A. {Brewster}, L.~M. {Ziurys}, and S.~{Wyckoff},
  \emph{\apj} \textbf{634}, 1126--1132 (2005).

\bibitem[{McClure-Griffiths} et~al.(2005)]{McClure+2005}
N.~M. {McClure-Griffiths}, J.~{Dickey}, B.~M. {Gaensler}, A.~J. {Green},
  M.~{Haverkorn}, and S.~S., \emph{\apjs} \textbf{158}, 178--187 (2005).

\bibitem[{Pandian} et~al.(2008)]{pandian+2008}
J.~{Pandian}, E.~{Momjian}, and P.~{Goldsmith}, \emph{\aap} \textbf{486},
  191--208 (2008).

\bibitem[{Weiss et al.}(2010)]{weiss}
{Weiss et al.}, \emph{\aap} \textbf{in press} (2010).

\bibitem[{Van der Werf et al.}(2010)]{vdwerf}
{Van der Werf et al.}, \emph{A\&A} \textbf{518}, L42 (2010).

\bibitem[{Lis} et~al.(2010)]{Lis2010b}
D.~C. {Lis}, J.~C. {Pearson}, and D.~A. {Neufeld et al.}, \emph{\aap\ in press,
  preprint arXiv:1007.1461}  (2010).

\bibitem[{Schilke} et~al.(2010)]{Schilke+2010}
P.~{Schilke}, C.~{Comito}, and H.~S.~P. {M\"uller et al.}, \emph{\aap\ in
  press, preprint arXiv:1007:0670}  (2010).

\bibitem[{Hily-Blant et al.}(2010)]{hilyblant2010}
{Hily-Blant et al.}, \emph{\aap} \textbf{in press} (2010).

\bibitem[{van Dishoeck} et~al.(1993)]{vanDishoeck1993}
E.~F. {van Dishoeck}, D.~J. {Jansen}, P.~{Schilke}, and T.~G. {Phillips},
  \emph{\apjl} \textbf{416}, L83--86 (1993).

\end{thebibliography}

\IfFileExists{\jobname.bbl}{}
 {\typeout{}
  \typeout{******************************************}
  \typeout{** Please run "bibtex \jobname" to optain}
  \typeout{** the bibliography and then re-run LaTeX}
  \typeout{** twice to fix the references!}
  \typeout{******************************************}
  \typeout{}
 }

\end{document}